\documentclass{PoS}

\usepackage{mathrsfs}
\usepackage{amsmath}
\usepackage{slashed}

\title{Inhomogeneous condensates in the parity doublet model}

\ShortTitle{CDW in the parity doublet model}

\author{\speaker{Achim Heinz}\\ 
        Institut f\"{u}r Theoretische Physik\\
	Johann Wolfgang Goethe-Universit\"{a}t\\
	Max-von-Laue-Str. 1 \\
	60438 Frankfurt am Main, Germany\\

        E-mail: \email{heinz@th.physik.uni-frankfurt.de}}

\abstract{Within the parity doublet model coupled to the linear sigma model including vector mesons
it is possible to describe vacuum properties of the low energy mesons and to achieve nuclear
saturation at nonzero density. Motivated by recent studies we investigate the emergence of
inhomogeneous condensation in the parity doublet model. As a first step the chiral density wave
(CDW) is considered, which allows for a straightforward investigation of inhomogeneous
condensation. As a result it is still possible to have a homogeneous ground state of nuclear matter,
but at larger baryon chemical potential the CDW is favored with respect to the homogeneous phase. }

\FullConference{Xth Quark Confinement and the Hadron Spectrum,\\
		October 8-12, 2012\\
		TUM Campus Garching, Munich, Germany}

\begin{document}

\section{Introduction}

For many decades the phase diagram of the quantum chromodynamics (QCD) was believed to be simple
\cite{Cabibbo:1975ig}. Only two phases were considered: a confined phase, where the relevant degrees
of freedom are hadrons which are build from quarks and gluons, and a deconfined phase, where the
relevant degrees of freedom are quarks and gluons \cite{Rischke:2003mt}. 

With increasing computational power, lattice simulations were able to test the above mentioned
picture at nonzero temperature and zero quark chemical potential \cite{Aoki:2006br,Cheng:2006qk}.
Since, at present lattice simulations are not able to access the high quark potential regime,
effective models and QCD-like theories have to be used. These approaches suggest a different
picture, according to which a rich phase diagram with a complicated structure is realized. 
At low temperatures and at low densities matter is confined and the relevant degrees of freedom are
hadrons; on the contrary, at very high densities a color superconducting phase should be present
\cite{Rischke:2003mt}, but it is unclear how the transition from nuclear matter to such phases looks
like. Also the critical quark chemical potential for a transition to such a phase is unknown.
Beside these uncertainties, there is also the possibility that further states of matter e.g.
quarkyonic matter \cite{Kojo:2009ha} exist.

All approaches to describe QCD at finite density have to use simplifications. For example the NJL
model \cite{Klevansky:1992qe} takes only quark degrees of freedom into account and therefore is not
the best choice to describe the regime of confined matter. On the other hand, models based on
hadrons have no chances to access the regime where quarks and gluons are the relevant degrees of
freedom. 

The general idea of inhomogeneous condensation goes back to Migdal \cite{Migdal:1978az}. He first
introduced inhomogeneous condensation within nuclear matter which was realized via the famous
chiral density wave (CDW). The idea of inhomogeneous condensation never got out of fashion and only
a few years ago the Gross-Neveu model in two dimensions could be solved exactly, analytically
\cite{Schnetz:2004vr} and numerically \cite{Wagner:2007he}. It has be shown that for high densities
and low temperatures inhomogeneous condensation dominates the phase diagram. Based on these results
it was shown that the NJL-model in four dimensions shows the same class of modulations as
the Gross-Neveu model \cite{Nickel:2009wj}.

It is of major interest to also look for the possibility of inhomogeneous condensation within a
fully hadronized model that is able to describe vacuum properties and to describe nuclear matter
saturation. The parity doublet model \cite{Jido:1998av} successfully reproduces vacuum
phenomenology \cite{Gallas:2009qp} and physics at finite density \cite{Zschiesche:2006zj}. The next
step, achieved here, is to test the ground state at nonzero density for the formation of a
inhomogeneous condensation. Furthermore a generalization of the parity doublet model is
straightforward and also exotic matter states like tetraquarks and glueballs can be introduced
\cite{Gallas:2011qp}. In this work however, we use the model studied in Ref.\
\cite{Zschiesche:2006zj}.  

\section{The Model}

The mesonic part of the Lagrangian is a $SU(2)$ linear sigma model including vector mesons
\cite{Walecka:1974qa}. It has the following form:
\begin{align}
 \mathscr{L}_M &= \frac{1}{2}\partial_\mu \sigma \partial^\mu \sigma + \frac{1}{2} \partial_\mu
\vec{\pi} \partial^\mu \vec{\pi}
-\frac{1}{4} F_{\mu \nu}F^{\mu \nu} \nonumber \\ 
&+ \frac{1}{2} m^2 (\sigma^2 + \vec{\pi}^2) + \frac{1}{2}m_\omega^2 \omega_\mu \omega^\mu -
\frac{\lambda}{4}(\sigma^2 
+ \vec{\pi}^2)^2 + \epsilon \sigma ~, 
\end{align}
with the field strength tensor for the vector meson fields $F_{\mu \nu} = \partial_\mu
\omega_\nu-\partial_\nu \omega_\mu$. In the vacuum chiral symmetry is spontaneous broken; this is
achieved in the model via the sign of the parameter $m^2$. The vacuum exception value (v.e.v.) of
the field $\sigma$ is nonzero and its value $\sigma_0 = \varphi$ corresponds at zero temperature
and density to the pion decay constant, $\varphi = f_\pi$.

The baryon part of model considers, besides the nucleon $N$, also its chiral partner $N^*$ 
\cite{Zschiesche:2006zj}. The Lagrangian is formulated in terms of the bare fields $\psi_1$ and
$\psi_2$: These fields are chiral eigenstates but not mass eigenstates of the model. The physical
fields $N$ and $N^*$ emerge as superpositions of $\psi_1$ and $\psi_2$. The fields $\psi_1$ and
$\psi_2$ transform under $SU_L(2) \times SU(2)_R$ the following way:
\begin{align}
\psi_{1, R} \rightarrow U_R ~\psi_{1, R} ~, ~~~~ \psi_{1, L} \rightarrow U_L ~\psi_{1, L} ~,
~~~~~~~~~
\psi_{2, R} \rightarrow U_L ~\psi_{2, R} ~, ~~~~ \psi_{2, L} \rightarrow U_R ~\psi_{2, L} ~. 
\end{align}
Notice that $\psi_2$ transforms in a mirror way with respect to the field $\psi_1$. Besides the well
known kinetic terms and the coupling to mesons, the mirror assignment allows to construct a further
chiral invariant bilinear term:
\begin{align}
 \bar{\psi}_{2, L} \psi_{1, R} - \bar{\psi}_{2, R} \psi_{1, L} - \bar{\psi}_{1, L}
\psi_{2, R} +
\bar{\psi}_{1, R} \psi_{2, L} 
 = \bar{\psi}_2 \gamma_5 \psi_1 - \bar{\psi}_1 \gamma_5 \psi_2  ~.
\end{align}
The full Lagrangian in the parity doublet model is:
\begin{align}
 \mathscr{L}_B &= \bar{\psi}_{1} \imath \slashed \partial \psi_{1} -\frac{1}{2} \hat{g}_1 
\bar{\psi}_{1}
\left( \sigma + \imath \gamma_5 \vec{\tau} \cdot \vec{\pi} \right) \psi_{1} 
               + \bar{\psi}_{2} \imath \slashed \partial \psi_{2} - \frac{1}{2} \hat{g}_2 
\bar{\psi}_{2} 
\left( \sigma - \imath \gamma_5 \vec{\tau} \cdot \vec{\pi} \right) \psi_{2} \nonumber \\
&- g_\omega^{(1)}\bar{\psi}_{1} \imath \gamma_0 \omega_0 \psi_{1} 
- g_\omega^{(2)}\bar{\psi}_{2} \imath \gamma_0 \omega_0 \psi_{2}   
+ m_0 \left( \bar{\psi}_{2} \gamma_5 \psi_{1} - \bar{\psi}_{1} \gamma_5 \psi_{2} \right)
+\mathscr{L}_M ~.
\end{align}
The term proportional to $m_0$ generates an additional contribution to the mass of the nucleons as
well as a mixing between $\psi_1$ and $\psi_2$. This leads to the aforementioned difference between
the chiral and mass eigenstates. In fact in order to obtain the physical masses of the nucleons,
$\psi_1$ and $\psi_2$ have to be transformed according to the following transformations:
\begin{align}
 \left(\begin{array}{c} 
  N^* \\
  N \\
\end{array} \right)  = \frac{1}{\sqrt{2 \text{cosh} \delta } }
\left(\begin{array}{cc} 
  \exp(\delta/2) & \gamma_5 \exp(-\delta/2) \\
\gamma_5 \exp(-\delta/2) & -\exp(\delta/2) \\
\end{array} \right) 
 \left(\begin{array}{c} 
  \psi_1 \\
  \psi_2 \\
\end{array} \right) ~.
\end{align}
Resulting from this transformation the now physical masses of the $N^*$ and $N$ resonances read:
\begin{align}
 m_N &= 
\frac{1}{2}\sqrt{\left( \frac{1}{2}\hat{g}_1 + \frac{1}{2}\hat{g}_2\right)^2 \varphi^2 + 4 m_0^2} 
+ \frac{1}{2}\left( \frac{1}{2}\hat{g}_1 - \frac{1}{2}\hat{g}_2\right) \varphi ~, \nonumber\\
m_{N^*} &= 
\frac{1}{2}\sqrt{\left( \frac{1}{2}\hat{g}_1 + \frac{1}{2}\hat{g}_2\right)^2 \varphi^2 + 4 m_0^2}
- \frac{1}{2}\left( \frac{1}{2}\hat{g}_1 - \frac{1}{2}\hat{g}_2\right) \varphi ~.
\end{align}
From the equations for the nucleon masses, the relevance of the parameter $m_0$ is clear. It allows
for a mass of the nucleon even if the chiral symmetry is restored, $\varphi = 0$. The chiral
condensate is not solely responsible for the mass of the baryons, but generates the mass splitting
between the nucleon and its chiral partner. Sending the parameter $m_0 \rightarrow 0$ one reobtains
the naive assignment, and the masses are generated only by the chiral condensate. 

The parity doublet model combines the linear sigma model with the nucleon $N$ and it's chiral
partner $N^*$. In the mean field approximation \cite{Serot:1984ey} the Lagrangian reduces to
the potential:
\begin{align}
 \mathscr{V}_M = - \frac{1}{2} m^2 \varphi^2 - \frac{1}{2}m_\omega^2 \bar{\omega}_0^2 +
\frac{\lambda}{4}\varphi^4 - \epsilon \varphi~,
\end{align}
and the grand canonical potential in the no see approximation reads:
\begin{align}
 \Omega/V = \mathscr{V}_M + \sum_j \frac{\gamma_j}{(2 \pi)^3} 
\int_{-\infty}^\infty d^3k \left( E_j(k) - \mu_j^* \right) \theta \left(E_j(k) - \mu_j^* \right)~,
\end{align}
where the sum runs over the resonances $N$ and $N^*$ with the corresponding fermionic degeneracy
factor $\gamma_j$. The energy has the known form $E_j = \sqrt{k^2 + m_j^2}$. The $\theta$ function
requires the chemical potential to be $\mu_j^* = \mu_j - g_\omega \bar{\omega}_0 = \sqrt{k^2 +
m_j^2}$. Minimizing $\Omega$ with respect to $\varphi$ and $\bar{\omega}_0$ leads to the nonzero
density behavior. This model has been studied and extended in Refs.\
\cite{Gallas:2009qp,Gallas:2011qp}. 

Following the promising results in the above mentioned publications, the model has to be tested for
inhomogeneous condensation. A straightforward approach is the CDW, which can be parameterized in the
following way \cite{Migdal:1978az}:
\begin{align}
 \left\langle \sigma \right\rangle = \varphi \cos (2 f x) ~, ~~~~ 
 \left\langle \pi_0  \right\rangle = \varphi \sin (2 f x) ~, ~~~ \text{where $x$ is a spatial
coordinate.} 
\end{align}
In the limit $f \rightarrow 0$ the v.e.v. is a constant equal to $\varphi$, and the ordinary parity
doublet model is realized. Inserting this Ansatz the Lagrangian for the baryons as well as for the
mesons changes. For the potential $\mathscr{V}_M$ a further contribution $2 f^2 \varphi^2$ follows.
The contribution arises from the kinetic term for the pions and sigma. As expected, this term
suppresses the formation of a CDW. Due to the fact that the CDW is suppressed in the vacuum the
parameters remain the same as in the case without CDW. Also the explicit symmetry breaking term get
slightly modified at finite baryon chemical potential.      

For the baryons the modifications are more demanding. The Lagrangian $\mathscr{L}_B$ has now a
explicit coordinate space dependency. By transforming the fields $\psi_1$ and $\psi_2$
as \cite{Ebert:2011rg}:
\begin{align}
 \bar{\psi}_{1} &\rightarrow \bar{\psi}_{1} \exp[-\imath \gamma_5 \tau_3 f x] ~, 
 \psi_1         \rightarrow \exp[-\imath \gamma_5 \tau_3 f x] \psi_1  ~, \\
 \bar{\psi}_{2} &\rightarrow \bar{\psi}_{2} \exp[+\imath \gamma_5 \tau_3 f x] ~,
 \psi_2         \rightarrow \exp[+\imath \gamma_5 \tau_3 f x] \psi_2  ~,
\end{align}
the Lagrangian $\mathscr{L}_B$ takes the following form:
\begin{align}
 \mathscr{L}_B 
=& \bar{\psi}_{1} \imath \slashed \partial  \psi_{1} 
+  \bar{\psi}_{1} \gamma_1 \gamma_5 \tau_3 f\psi_{1} 
+  \bar{\psi}_{2} \imath \slashed \partial  \psi_{2} 
-  \bar{\psi}_{2} \gamma_1 \gamma_5 \tau_3 f\psi_{2} 
\nonumber \\
&- \frac{1}{2} \hat{g}_1 \varphi \bar{\psi}_{1} \psi_{1} - \frac{1}{2} \hat{g}_2 \varphi \bar{\psi}_{2} \psi_{2} 
+ m_0 \left( \bar{\psi}_{2} \gamma_5 \psi_{1} -\bar{\psi}_{1} \gamma_5 \psi_{2} \right)~.
\end{align}
For illustrative purpose the additional $\bar{\omega}_0$ dependency and baryon chemical potential
$\mu$ is not shown, but it is simple to generalize the expressions. The diagonalization of the
Lagrangian leads to a splitting of the energy eigenstates. In the presence of the CDW four different
energy eigenstates emerge, in contrary to the homogeneous case where only $N$ and $N^*$ are present.
The different states can be written down in the general form $E_k(p) = \sqrt{p^2 +
\bar{m}_k(p_1)^2}$, $k = 1\dots4$, where $\bar{m}_k(p_1)^2$ is a momentum dependent mass which has
to be calculated numerically. The grand canonical potential is given by:
\begin{align}
\Omega/V 
&= 2f^2\varphi^2 +\frac{1}{4}\lambda\varphi^4 -\frac{1}{2}m^2\varphi^2- \epsilon\varphi
-\frac{1}{2}m_\omega^2\bar{\omega}_0^2 \nonumber \\
&+\sum_{k=1}^4 \frac{2}{(2 \pi)^2} \int_{-\infty}^\infty dp_1~  
 \Theta \left[\mu^*- E_k(p_1)\right]  \frac{1}{6} \left[3 \mu^* E_k(p_1)^2 - (\mu^*)^3 -2 E_k(p_1)^3
\right] ~,
\end{align}
with $\mu^* = \mu-g_\omega \bar{\omega}_0$. The behavior at nonzero baryon chemical potential can be
found by extremizing the grand canonical potential with respect to the dynamical degrees of
freedom:
\begin{align}
 0 \stackrel{!}{=} \frac{\partial (\Omega/V)}{\partial \bar{\omega}_0} ~, ~~
 0 \stackrel{!}{=} \frac{\partial (\Omega/V)}{\partial \varphi} ~,~~
 0 \stackrel{!}{=} \frac{\partial (\Omega/V)}{\partial f} ~.
\end{align}

The parameters $g_\omega$, $\lambda$, $m$, $\epsilon$ and $m_\omega$ are fixed using the known
vacuum values for the masses of the pions and omega, as well the pion decay constant ($2 \lambda
\varphi^2 = m_\sigma^2 - m_\pi^2$, $2 m^2 = m_\sigma^2 - 3 m_\pi^2$, $\epsilon = m_\pi^2 f_\pi $)
and the conditions for nuclear matter saturation:
\begin{align}
 \frac{E}{A} - m_N = -16 ~\text{MeV},~~\text{and} ~~\rho_0 = 0.16 ~\frac{1}{\text{fm}^3} ~.
\end{align}
The first equation ensures a binding energy per nucleon and the second the density at the minimum
(it corresponds to the baryon chemical potential of $\mu = 923 ~\text{MeV}$). The parameter $m_0$
remains as a input parameter. Using in addition the masses for $N$ and $N^*$ the parameters
$\hat{g}_1$ and $\hat{g}_2$ can be fixed.

\section{Results}

The phase diagram of the parity double model shows a rich structure at dense hadronic matter. Since
the model is based on Ref.\ \cite{Zschiesche:2006zj} we use their parameters as a starting
point. For different values of the mass $m_{N^*}$ and different values for the parameter $m_0$ the
chiral condensate $\varphi$ is calculated as a function of the baryon chemical potential $\mu$. A
feature of the model is to achieve nuclear saturation. For all combinations of the values for
$m_{N^*}$ and $m_0$ the parameters $m$ and $g_\omega$ are tuned in a way to fullfill the
conditions for nuclear saturation at a baryon chemical potential of $\mu = 923 ~\text{MeV}$. The
results are shown in Fig.\ 1. On the left hand side $m_{N^*}$ is low, $m_{N^*} = 1200
~\text{MeV}$ and on the right hand side is high, $m_{N^*} = 1500 ~\text{MeV}$. For both cases the
parameter $m_0$ is varied from $600 ~\text{MeV}$ to $700 ~\text{MeV}$ and finally to $800
~\text{MeV}$. A general result for all parameter sets for the behavior of $\varphi$ is the
presence of three different phases. 

For small $\mu$ the chiral symmetry is broken and the value for the condensate $\varphi$ remains
constant at the vacuum value $f_\pi$. For an intermediate range for $\mu$ in the order of $\mu = 923
~\text{MeV}$ one obtains a first order phase transition. The value for the condensate $\varphi$
drops from the vacuum value $f_\pi$ to $30 - 60 ~\text{MeV}$, where the ground state is
still homogeneous and the CDW is not favored. In this range nuclear matter is formed. It shows that
for the formation of homogeneous nuclear matter the mass of the pion plays a crucial role. Sending
$m_\pi \rightarrow 0$, the CDW seems to be always the favored ground state. This is unaffected by
the choice of $m_0$. For large $\mu$ a second first order phase transition occurs. The CDW is now
always the favored ground state. Also for asymptotic large values for $\mu$ chiral symmetry will
not be restored and the CDW remains the favored state of matter.

For $m_{N^*} = 1500 ~\text{MeV}$ (left panel of Fig.\ 1) the second phase transition shows a
strong dependence on the parameter $m_0$. For small values of $m_0$ the transition happens at large
$\mu$ and for a larger $m_0$ the intermediate homogeneous ground state shrinks to a small range of
$\mu$. The value for $\varphi$ after the second phase transition also displays a dependence on
$m_0$, still the overall behavior is the same. 

For $m_{N^*} = 1200 ~\text{MeV}$ (right panel of Fig.\ 1) the situation is slightly different: the
behavior of $\varphi$ shows no $m_0$ dependency after the second phase transition. Furthermore for
$m_0 = 600 ~\text{MeV}$ and $m_0 = 700 ~\text{MeV}$ the critical $\mu$ is almost the same.

\begin{figure}
\includegraphics[scale=0.58] {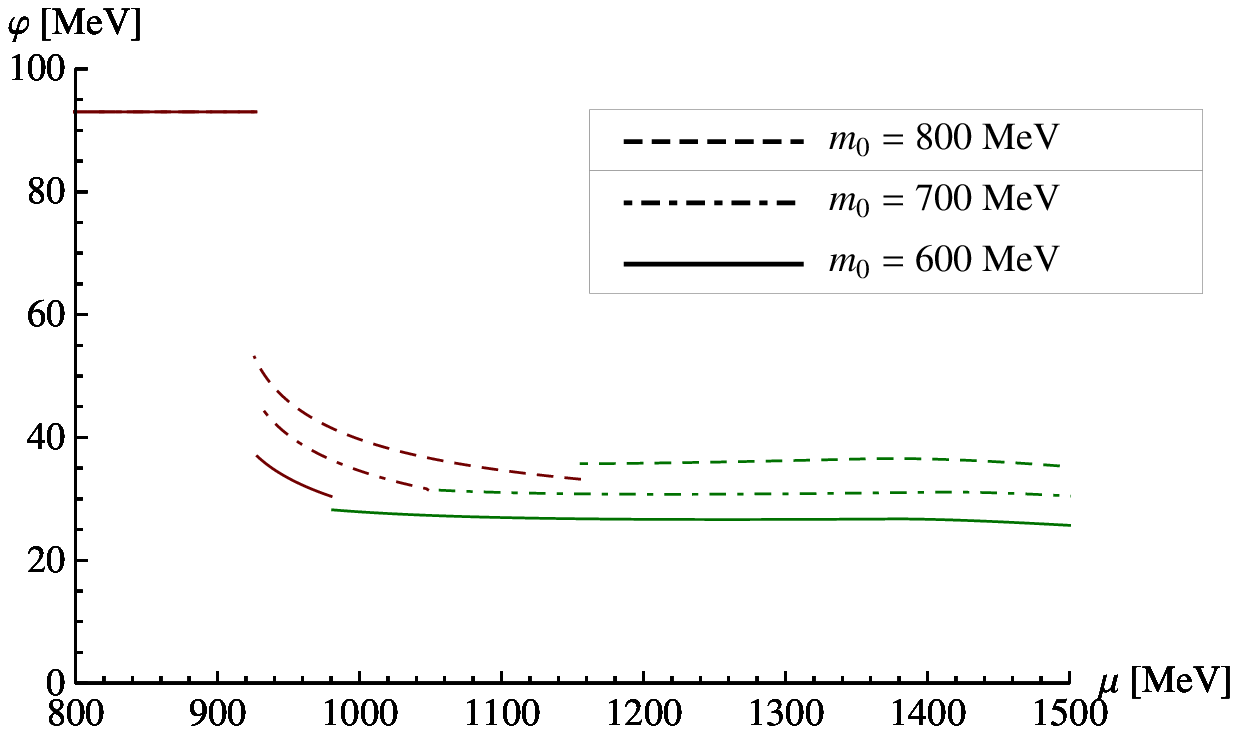}
\includegraphics[scale=0.58] {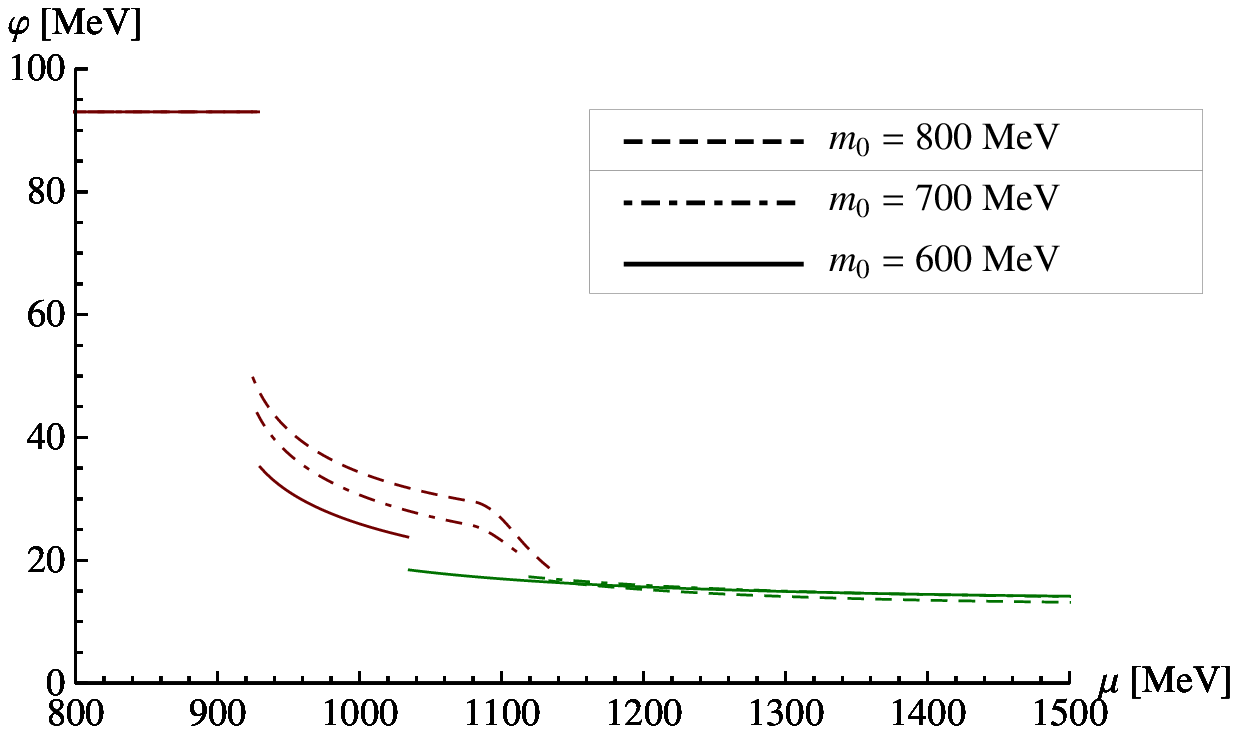}
\caption{The chiral condensate $\varphi$ as a function of the baryon chemical potential $\mu$ for
different values of $m_0$ and $m_{N^*}$. On the left panel with a large mass for the
chiral partner of the nucleon $m_{N^*}=1500 ~\text{MeV}$ and on the right panel a small mass
$m_{N^*}=1200 ~\text{MeV}$. The red color indicates the ground state to be homogeneous, while the
green that the CDW is favored.}%
\label{sigma_of_mu}
\end{figure}
Another interesting point is the dispersion relation in the inhomogeneous phase, see Fig. 2
left panel. It shows the that the spectrum splits from two energy levels in the homogeneous phase
to four in the inhomogeneous one. Moreover, for the lower two the energy decreases with increasing
momentum and the upper two increase with increasing momentum. The parameters are the same as in the
case for $m_0 = 800 ~\text{MeV}$ and $\mu$ is slightly larger than the critical value for the
transition to the CDW. Using the same parameters as in the left plot of Fig.\ 2 the relative
densities are shown. Within the homogeneous region the transition to nuclear matter is found.

\begin{figure}
\includegraphics[scale=0.69] {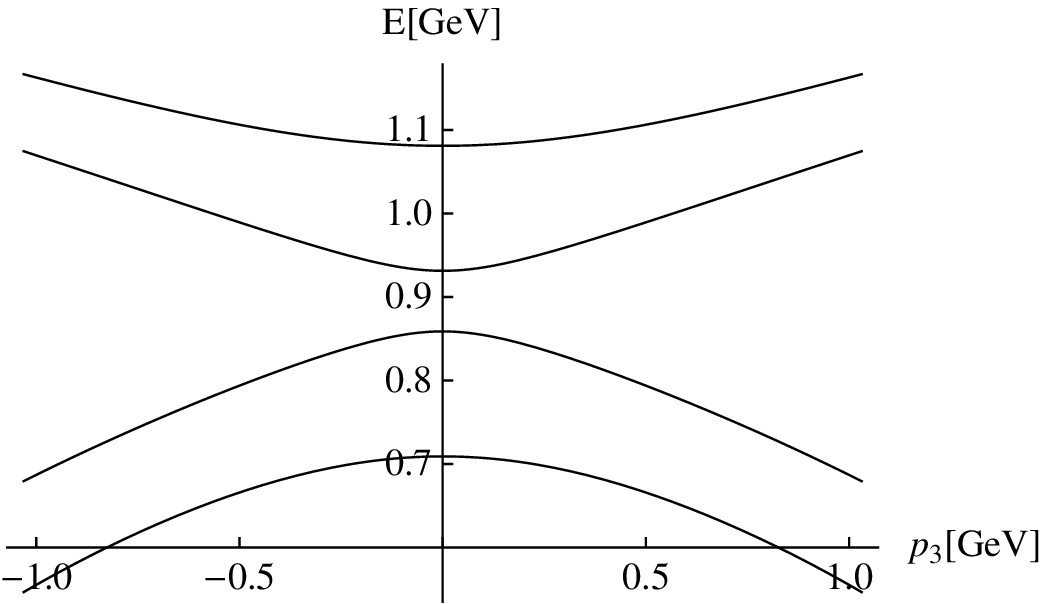}
\includegraphics[scale=0.69] {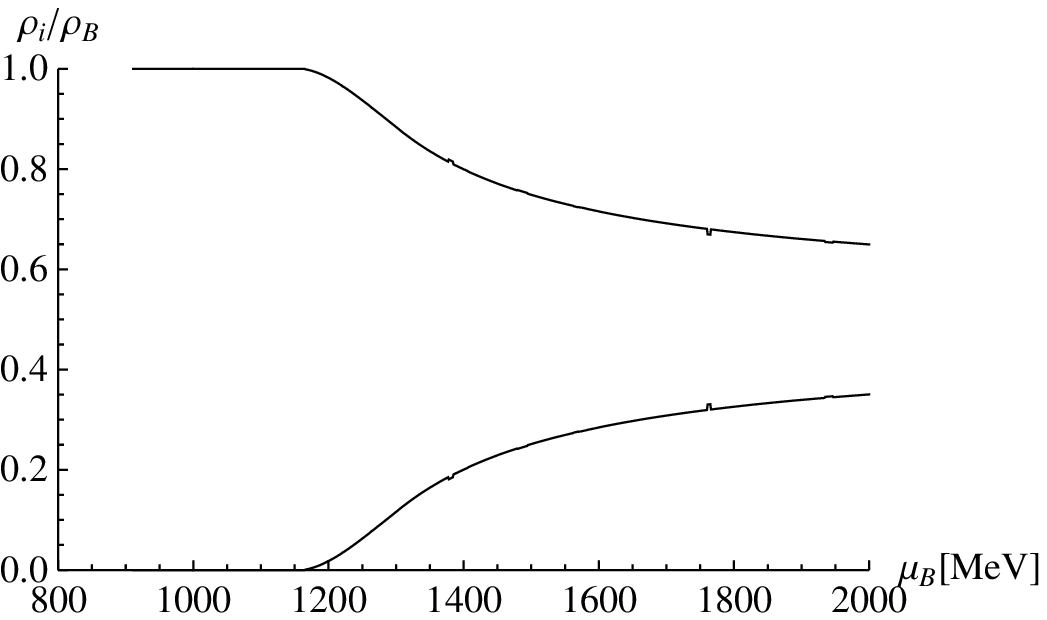}
\caption{The left plot shows the dispersion relation for the CDW. The spectrum shows four
different energy levels in contrast to the homogeneous ones. The right shows the relative density of
the states at finite $\mu$.}%
\label{dis_rel}
\end{figure}

\section{Summary}

Inhomogeneous condensation is a relevant feature of the parity doublet model coupled to the linear
sigma model with vector mesons. In a first step it has been shown that for $m_\pi = 139
~\text{MeV}$ homogeneous nuclear matter is obtained. At a large baryon chemical potential the
CDW is the favored state of matter. Sending the pion mass to zero (chiral limit) the picture
changes. It is now no longer possible to find a homogeneous ground state. Furthermore, in both cases
chiral symmetry does not get restored at high baryon chemical potential and the chiral condensate
even increases for increasing density. 

For the future the model has to be further extended. The CDW is just one of many different
realizations of inhomogeneous condensation. A more general ansatz should be used
\cite{Schnetz:2004vr}. The extended linear sigma model (eL$\sigma$m)
\cite{Parganlija:2012fy} has proven to be robust in describing vacuum
phenomenology. Therefore combining both the parity doublet and the eL$\sigma$m would allow to
constrain the existents of CDW. Also the effect to the onset of the CDW regime of exotic matter like
tetraquarks or glueballs \cite{Gallas:2011qp} has to be
tested in a more general framework.
   
\section{Acknowledgments}
The authors acknowledges support from the Helmholtz Research School on Quark Matter Studies and
thanks D. H. Rischke, M. Wagner and F. Giacosa for valuable discussion.

\end{document}